\documentclass[12pt]{iopart}

\usepackage{bm}%
\usepackage{graphics}%
\usepackage{graphicx}%
\usepackage{epsfig}%
\usepackage{subfig}
\usepackage{multirow}
\usepackage{float}
\usepackage{dcolumn}
\usepackage{upgreek}
\usepackage{cite}
\usepackage{setspace}
\usepackage{color}
\usepackage[outdir=./]{epstopdf}
\begin{document}

\title[]{Active control of electromagnetically induced transparency analogue in all-dielectric metamaterials loaded with graphene}

\author{Tingting Liu$^{1}$, Huaixing Wang$^{1}$, Chaobiao Zhou$^{2}$, Xiaoyun Jiang$^{3}$ and Shuyuan Xiao$^{4,5}$}

\address{$^{1}$~Laboratory of Millimeter Wave and Terahertz Technology, School of Physics and Electronics Information, Hubei University of Education, Wuhan 430205, People's Republic of China}
\address{$^{2}$~College of Mechanical and Electronic Engineering, Guizhou Minzu University, Guiyang 550025, People's Republic of China}
\address{$^{3}$~Wuhan National Laboratory for Optoelectronics, Huazhong University of Science and Technology, Wuhan 430074, People's Republic of China}
\address{$^{4}$~Institute for Advanced Study, Nanchang University, Nanchang 330031, People's Republic of China}
\address{$^{5}$~Jiangxi Key Laboratory for Microscale Interdisciplinary Study, Nanchang University, Nanchang 330031, People's Republic of China}
\ead{syxiao@ncu.edu.cn}
\vspace{10pt}
\begin{indented}
\item[]August 2020
\end{indented}

\begin{abstract}
Electromagnetically induced transparency (EIT) analogue in all-dielectric metamaterials with a high quality factor provides an effective route to enhance light-matter interaction at the nanoscale. In particular, the active control applied to it enables great degree of freedom for spatial light modulation and thus promises functional device applications with high flexible tunability. Here we load graphene into all-dielectric metamaterials and realize the remarkably high modulation depth in the transmission amplitude of the EIT resonance with the manipulation of graphene conductivity, via shifting the Fermi level or altering the layer number. The physical origin lies in the controllable light absorption through the interband loss of graphene in the near infrared. This work reveals a strategically important interaction mechanism between graphene and EIT resonance in all-dielectric metamaterials, and open avenues in designing a family of hybrid metadevices that permit promising applications to light modulation, switching and ultrasensitive biosensing.
\end{abstract}

%
\vspace{2pc}
\noindent{\it Keywords}: electromagnetically induced transparency, optical modulation, all-dielectric metamaterials, graphene, near infrared
%
%
%
%

\section{Introduction}\label{sec1}

Metamaterials comprising periodically arranged subwavelength resonators have revolutionized the research field in manipulation of light-matter interaction due to the exotic properties inaccessible in natural materials\cite{zheludev2012metamaterials}. The performance of the conventional metal metamaterials is always affected by two kinds of losses, namely the radiative and nonradiative losses. Fortunately, the radiative loss can be suppressed by tailoring the structure geometry to enable the excitation of the so-called "trapped mode" weakly coupled to free space, or more ingeniously, to achieve the destructive interference using the near field coupling effect known as electromagnetically induced transparency (EIT) analogue\cite{fedotov2007sharp,zhang2008plasmon,liu2009plasmonic}. However, the nonradiative loss from the intrinsic resistance of metals, which is especially dominant in the near infrared and optical regimes, severely limits the quality (Q) factor up to 10 and hinders practical device applications. Therefore, it is necessary to seek low-loss materials as alternatives to metals for metamaterial resonators. Recently, exciting progress has been made on the issue of the nonradiative loss by adopting the Mie resonance of dielectric materials with high refractive index, such as silicon, germanium and tellurium\cite{jahani2016all,kuznetsov2016optically,Baranov2017}. In analogy to the excitation of the collective electrons at the metal surface, the oscillation of the displacement current in the dielectric resonators produces the magnetic dipole (first Mie resonance) and electric dipole (second Mie resonance) responses. This physical insight has inspired the excitation of the "trapped mode" and subsequently the implementation of the EIT resonance in all-dielectric metamaterials with extremely high Q factors up to a few hundreds, which shows great prospects for the realization of low-loss optical modulators, enhanced nonlinear sensors and slow light devices\cite{miroshnichenko2012fano,zhang2013near,zhang2014electromagnetically,yang2014all,hopkins2015interplay,liu2017high,tuz2018high,Li2019,Zhang2019}.

In recent years, the active control of optical responses is another research focus in the metamaterial field, which provides an additional degree of freedom to precisely manipulate the light-matter interaction. The realization approach in current works is through loading active materials into metamaterials and controlling the coupling effects between them under various external stimuli\cite{gu2012active,xu2016frequency,fan2017electromagnetic,manjappa2017hybrid,ahmadivand2017active,zhu2018controlling}. Particularly, graphene with unprecedented properties such as the dynamically tunable conductivity and ultrafast modulation response becomes an excellent candidate among all the active materials. So far, considerable efforts have been devoted to graphene metamaterials\cite{zhao2016graphene,he2017implementation,he2018graphene,xia2018plasmonically,Jia2019,Guan2020} or hybrid metal-graphene metamaterials for controllable EIT resonance\cite{li2016monolayer,xiao2017strong,chen2017study,xiao2018active,Hong2019,Xia2020}. Most recently, some pioneering works have paid attention to the coupling effect between graphene and the low-loss "trapped mode" in all-dielectric metamaterials\cite{argyropoulos2015enhanced,liu2017toroidal,Xiao2019,Sun2020}. However, the interaction between graphene and the classical EIT resonance in all-dielectric metamaterials in the near infrared has rarely to be explored.

In this work, we propose an active control of the EIT analogue in all-dielectric metamaterials through loading graphene into the unit cell. The numerical results show that the remarkably high modulation depth in the transmission amplitude of the EIT resonance can be realized with the manipulation of graphene conductivity, via shifting the Fermi level or altering the layer number, which is attributed to the controllable light absorption through the interband loss of graphene in the near infrared. The investigation on the interaction between graphene and the EIT resonance in all-dielectric metamaterials offers a great opportunity to compare with the modulation mechanism in the hybrid metal-graphene counterparts. Therefore, this work may inspire interest in developing a novel kind of active metadevice with functionalities attained through the exploitation of the controllable EIT resonance in the hybrid dielectric-graphene metamaterials.

\section{Geometric structure and numerical model}\label{sec2}

\begin{figure}[h]
\centering
\includegraphics[scale=0.60]{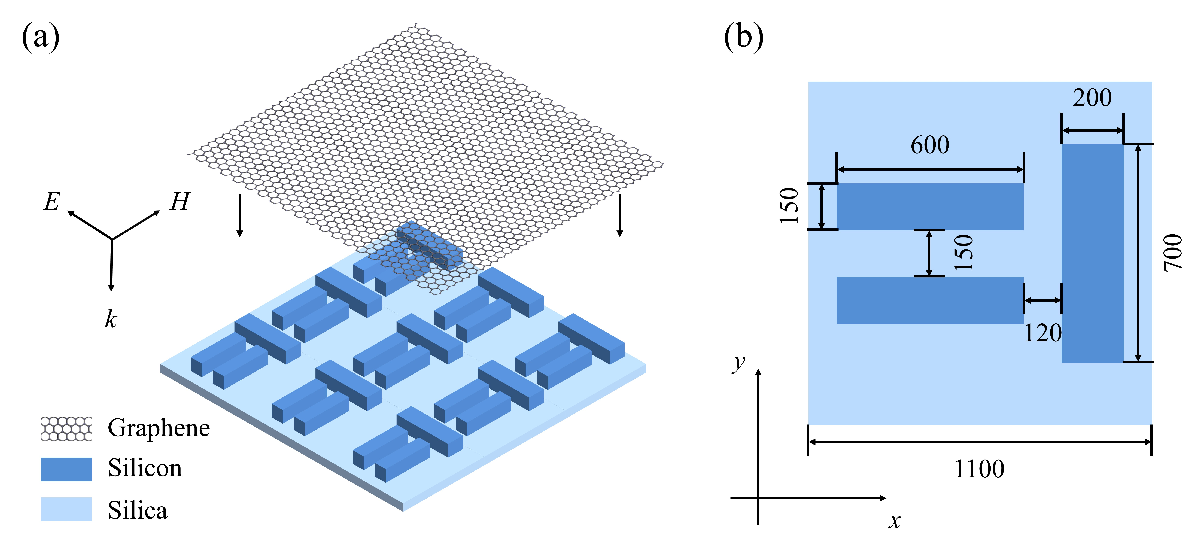}
\caption{(a) The schematic diagram of our proposed hybrid dielectric-graphene metamaterials. (b) The top view and the geometrical parameters of the unit cell.\label{fig1}}
\end{figure}

The schematic diagram and the geometrical parameters of the proposed hybrid dielectric-graphene metamaterials are illustrated in Figure. \ref{fig1}(a) and (b). The all-dielectric metamaterials employs a classical design composed of three silicon nanobars on a silica substrate for the EIT resonance, with the lattice constant of 1100 nm in the $x$ and $y$ directions. In the unit cell, the two parallel nanobars are both 600 nm in length and 150 nm in width, and are separated by 150 nm in the $y$ direction. The vertical nanobar is 700 nm in length and 200 nm in width, and is placed on the right side of the two parallel nanobars with a distance of 120 nm. All the nanobars are 160 nm in thickness. Then the monolayer graphene is transfered on the top of the all-dielectric metamaterials in a continuous morphology. The linearly plane wave is incident along the $-z$ direction with the electric field polarized in the $y$ direction.

\begin{figure}[h]
\centering
\includegraphics[scale=0.65]{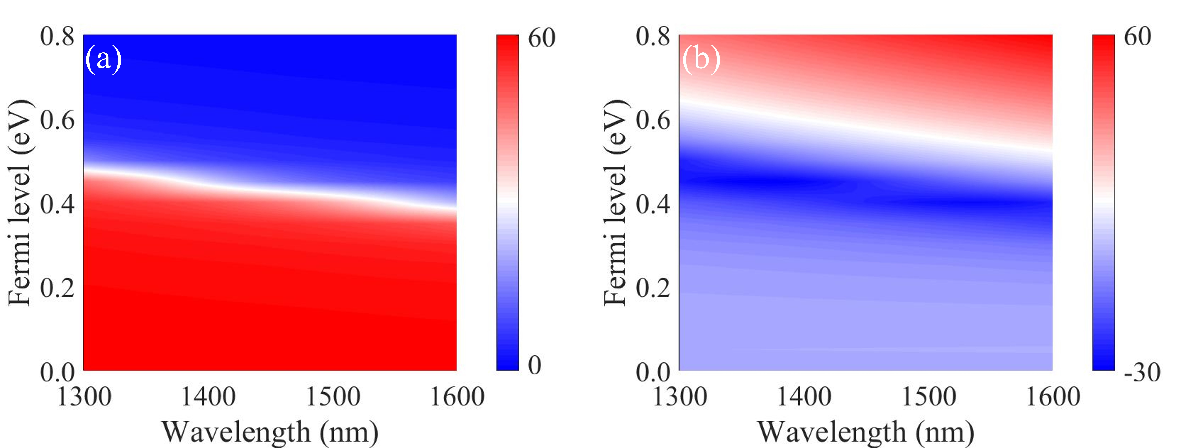}
\caption{(a) Real and (b) imaginary part of the wavelength dependent graphene conductivity as a function of the Fermi level.\label{fig2}}
\end{figure}

The full-wave numerical simulations are conducted using the finite-difference time-domain (FDTD) method. For the wavelength longer than 1200 nm in the near infrared, the silicon and the silica show extremely low loss and their constant refractive indices of $n_{\rm{Si}}=3.5$ and $n_{\rm{SiO}_{2}}=1.4$ are adopted for simulations\cite{palik1985handbook}. The monolayer graphene placed over the all-dielectric metamaterials is modeled as a 2D sheet and its conductivity can be derived from the random phase approximation (PRA) in the local limit, including the intraband and interband transitions\cite{zhang2015towards,xiao2016tunable},
\begin{eqnarray}
\sigma_{\rm{g}} =&\sigma_{\rm{intra}}+\sigma_{\rm{inter}}=\frac{2e^{2}k_{\rm{B}}T}{\pi\hbar^2}\frac{i}{\omega+i\tau^{-1}}\ln[2\cosh(\frac{E_{\rm{F}}}{2k_{\rm{B}}T})]  \\\nonumber
&+\frac{e^{2}}{4\hbar}[\frac{1}{2}+\frac{1}{\pi}\arctan(\frac{\hbar\omega-2E_{\rm{F}}}{2k_{\rm{B}}T})  \\\nonumber
&-\frac{i}{2\pi}\ln\frac{(\hbar\omega+2E_{\rm{F}})^{2}}{(\hbar\omega-2E_{\rm{F}})^{2}+4(k_{\rm{B}}T)^{2}}]\label{eq1},
\end{eqnarray}
where $e$ is the electron charge, $k_{\rm{B}}$ is the Boltzmann constant, $\hbar$ is the reduced Planck's constant, $\omega$ is the incident wave frequency, $E_{\rm{F}}$ is the Fermi level, the temperature $T$ is set to 300 K and the relaxation time $\tau=(\mu E_{\rm{F}})/(e v_{\rm{F}}^{2})$ is calculated from the carrier mobility $\mu=10000$ cm$^{2}$/V$\cdot$s and the Fermi velocity $v_{\rm{F}}=1\times 10^{6}$ m/s. As shown in Figure. \ref{fig2}(a) and (b), the wavelength dependent graphene conductivity can be continuously manipulated via shifting the Fermi level. Note that when the Fermi level is increased from the Dirac point by half of the photon energy, i.e., $E_{\rm{F}}>\hbar\omega/2$, the contribution of the interband transition is blocked due to the Pauli's exclusion principle. Therefore the real part of graphene conductivity dramatically decreases once the Fermi level exceeds the critical value, while the imaginary part continues to increase with significant contribution from the intraband transition.

\section{Simulation results and discussions}\label{sec3}

\begin{figure}[htbp]
\centering
\includegraphics[scale=0.80]{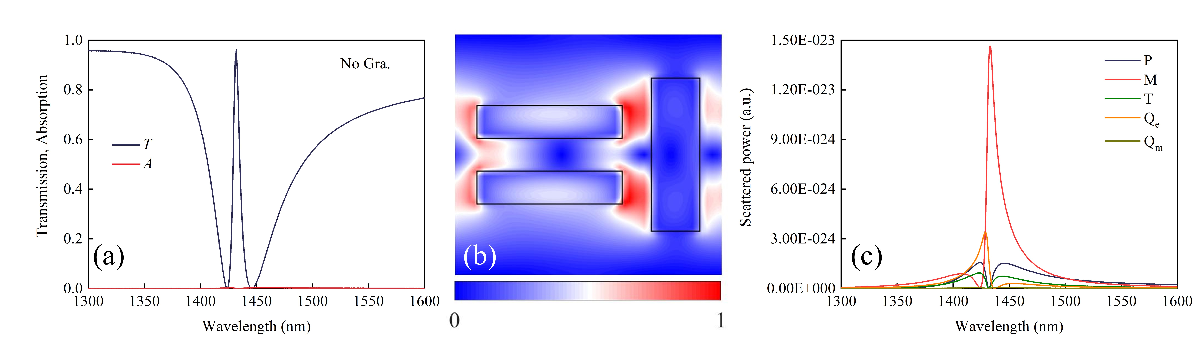}
\caption{(a)The simulated transmission and absorption spectra of the all-dielectric metamaterials without graphene. (b) Corresponding electric field distribution in the $x-y$ plane at resonance wavelength. (c) Scattered power of multipoles of the dielectric metasurface,including the electric dipole $P$, magnetic dipole $M$, toroidal dipole $T$, electric quadrupole $Q_{e}$ and magnetic quadrupole $Q_{m}$. \label{fig3}}
\end{figure}

We first investigate the EIT resonance in the all-dielectric metamaterials without the presence of graphene. Figure. \ref{fig3} (a) shows the simulated transmission and absorption spectra, we can see a sharp transparency window arising within a broad stop band at 1432 nm, exhibiting a typical EIT resonance behavior. The transmission amplitude is as high as 0.96. The corresponding Q factor can be calculated with the ratio of resonance wavelength of maximum transmission at the EIT window to the full width at half maximum of the resonance as $Q=\lambda_{\rm{res}}/\Delta\lambda$, with a value of 223, which is much higher than that in the conventional metal metamaterials. In the strucute, the incident light is firstly coupled into the vertical nanobar which serves as the bright resonator, and then the antisymmetric current oscillations are excited in the two parallel nanobars acting as a dark antenna via near field coupling with the vertical nanobar. Due to the close proximity, the destructive interference between these two excitation pathways conversely suppresses the radiation of the bright mode oscillation, leading to the EIT resonance with the confined electric field distributions in Figure. \ref{fig3} (b). The far-field scattered powers for different multipoles are calculated in a Cartesian coordinate system, as shown in  Figure. \ref{fig3} (c). In the vicinity of the resonance wavelength, the excitation of electric diople is greatly suppressed while the magnetic dipole becomes a dominant contributor to the resonance response, which provides a clear picture of the EIT resonance formation. In addition, note that the imaginary refractive indices of the dielectric materials, i.e., silicon and silica, are negligible in the concerned wavelength regime, the absorption loss in the all-metamaterials is also eliminated. Therefore, the EIT analogue with a high Q factor is demonstrated and the incident light is strongly trapped in the near field.

\begin{figure}[h]
\centering
\includegraphics[scale=0.45]{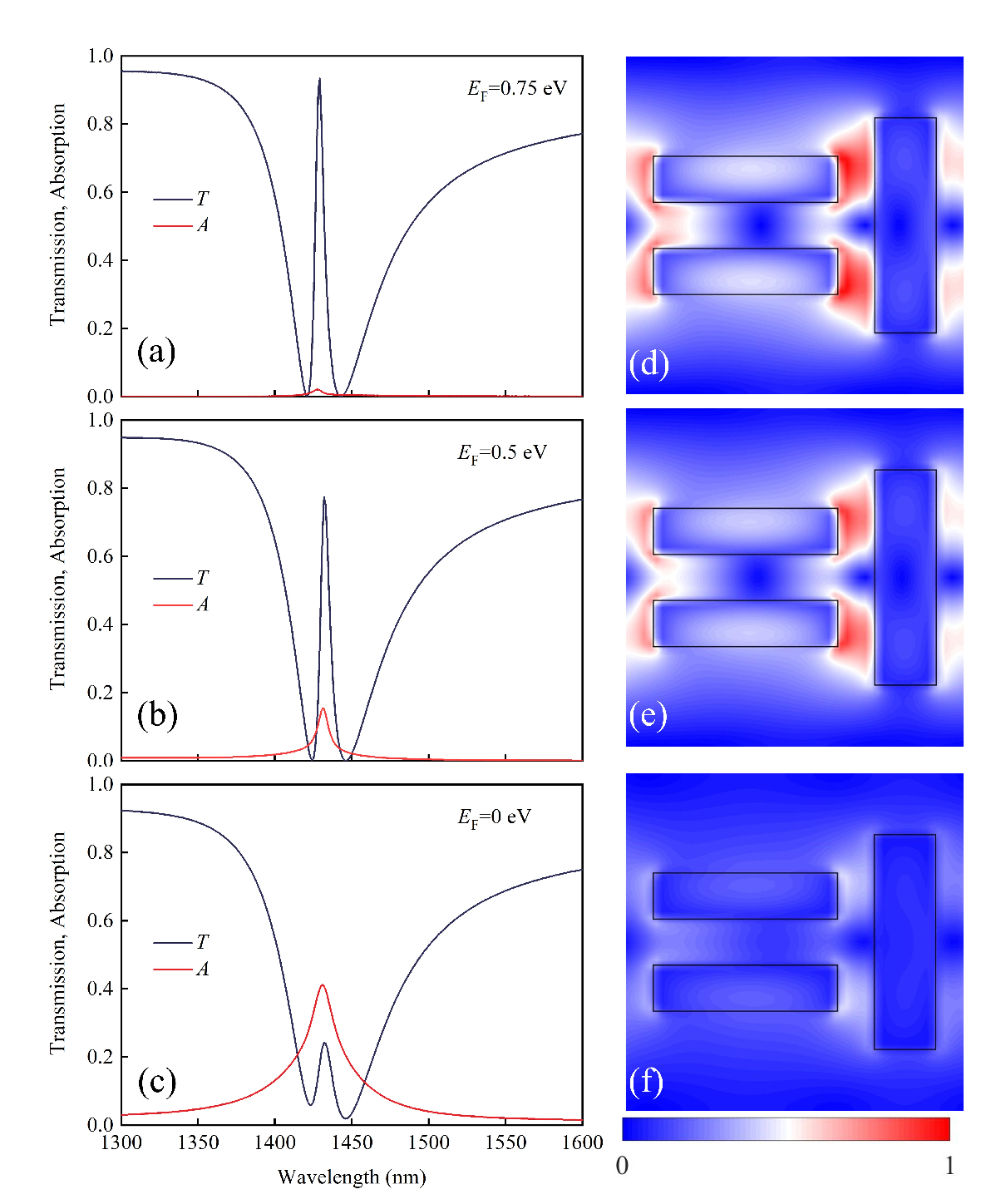}
\caption{(a)-(c) The simulated transmission and absorption spectra of the hybrid dielectric-graphene metamaterials via shifting the Fermi level. (d)-(f) Corresponding electric field profiles at EIT resonance. \label{fig4}}
\end{figure}

Next, we load the monolayer graphene into the all-dielectric metamaterials and investigate its modulation effect on the EIT resonance. Figure. \ref{fig4}(a)-(c) show the simulated transmission and absorption spectra, we can see that the EIT resonance undergoes a remarkable change in the transparency window via shifting the Fermi level of graphene, while the resonance wavelength remains nearly intact. When $E_{\rm{F}}$ initiates at 0.75 eV (heavily doped graphene), the transmission amplitude is 0.93, very close to the case without graphene, and the absorption peak of 0.02 begins to emerge at the resonance. When $E_{\rm{F}}$ gradually varies to 0.5 eV, the transmission declines to 0.77, and the absorption goes up to 0.15. When $E_{\rm{F}}$ finally reduces to 0 eV (undoped graphene), the transmission decreases to the minimum of 0.24, and the absorption reaches the maximum of 0.40. To quantitatively characterize the change in the transparency window, we introduce the modulation depth in the transmission amplitude as $\Delta T=(T_{\rm{0}}-T_{\rm{g}})\times 100\%$, where $T_{\rm{0}}$ and $T_{\rm{g}}$ refer to transmission amplitude at the transparency peak of the EIT analogue without and with the monolayer graphene, respectively. The modulation depth can be actively controlled via shifting the Fermi level, and the maximum $\Delta T=72\%$ is realized with the undoped graphene. It is noteworthy that EIT resonance position is insensitive to the varying Fermi level of graphene as the ultrathin graphene layer barely affects the resonance wavelength. 

The physics origin can be explained with the coupling effects between the all-dielectric metamaterials and the monolayer graphene. In Figure. \ref{fig4}(d)-(f), we plot the surface electric field distributions at the resonance wavelength. It is mentioned above that without the presence of graphene, the destructive interference gives rise to the EIT resonance and traps the incident light within the all-dielectric metamaterials, providing a local enhancement in the in-plane electric field. Here when graphene as the only lossy material in the hybrid structure is integrated, it strongly couples to incident light with its intrinsic absorption, leading to a remarkable change in the in-plane electric field and thus the resonance strength. With $E_{\rm{F}}=0.75$ eV (heavily doped graphene), the absorption just begins to emerge and the electric field remains nearly intact; with $E_{\rm{F}}=0.5$ eV, the absorption further goes up and the electric field slightly declines; with $E_{\rm{F}}=0$ eV (undoped graphene), the absorption reaches its maximum, corresponding to the fully degenerated electric field profile. Therefore, the modulation in the transmission is realized with the change in the resonance strength, which can be actively controlled with the light absorption of graphene. Note that in our previous works the on-to-off modulation of the EIT resonance in the THz hybrid metal-graphene metamaterials is realized with the increase of the Fermi level of graphene, showing an completely opposite variation tendency\cite{xiao2018active}. The key difference results from the fact that the real part of the graphene conductivity increases with the Fermi level in the THz regime while decreases with the Fermi level in the concerned near infrared. When the Fermi level decreases to half of the photo energy, i.e., 0.43 eV at the resonance wavelength of 1432 nm, the Pauli's blocking is relieved and the interband loss of graphene becomes dominate in the near infrared. As indicated in Figure. \ref{fig2}(a), the real part of the graphene conductivity dramatically increases with the decrease of Fermi level, which induces the light absorption enhancement and thus the resonance reduction.

\begin{figure}[htbp]
\centering
\includegraphics[scale=0.40]{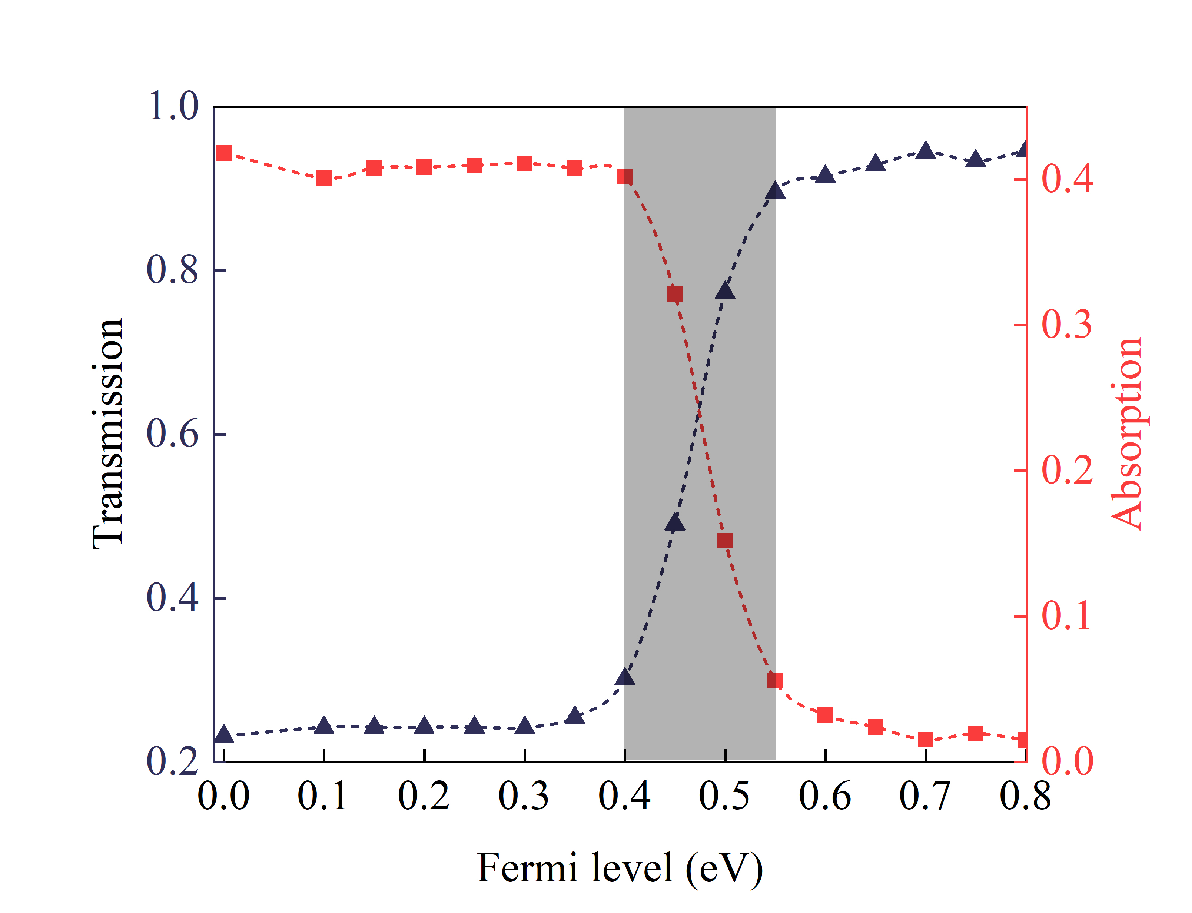}
\caption{The dependence of the transmission and absorption at resonance on the Fermi level of graphene.\label{fig5}}
\end{figure}

To gain a clearer insight, Figure. \ref{fig5} presents the dependence of the transmission and absorption at resonance on the Fermi level of graphene. Both amplitudes display dramatic changes around the critical Fermi level due to the Pauli's blocking effect, i.e., 0.43 eV and the transition region is denoted by the light grey area. The transmission is low when $E_{\rm{F}}<0.43$ eV and it increases rapidly as $E_{\rm{F}}$ goes through the critical point in transition region, reaching the saturation state at higher $E_{\rm{F}}$. During the process, the absorption of graphene undergoes the opposite change through the transition region, from the initial large values to the small values as $E_{\rm{F}}$ increases. Their exactly opposite variations further verify the physical origin of modulation of EIT resonance can be attributed to the controllable absorption through the interband loss of graphene via shifting the Fermi level.  

\begin{figure}[h]
\centering
\includegraphics[scale=0.40]{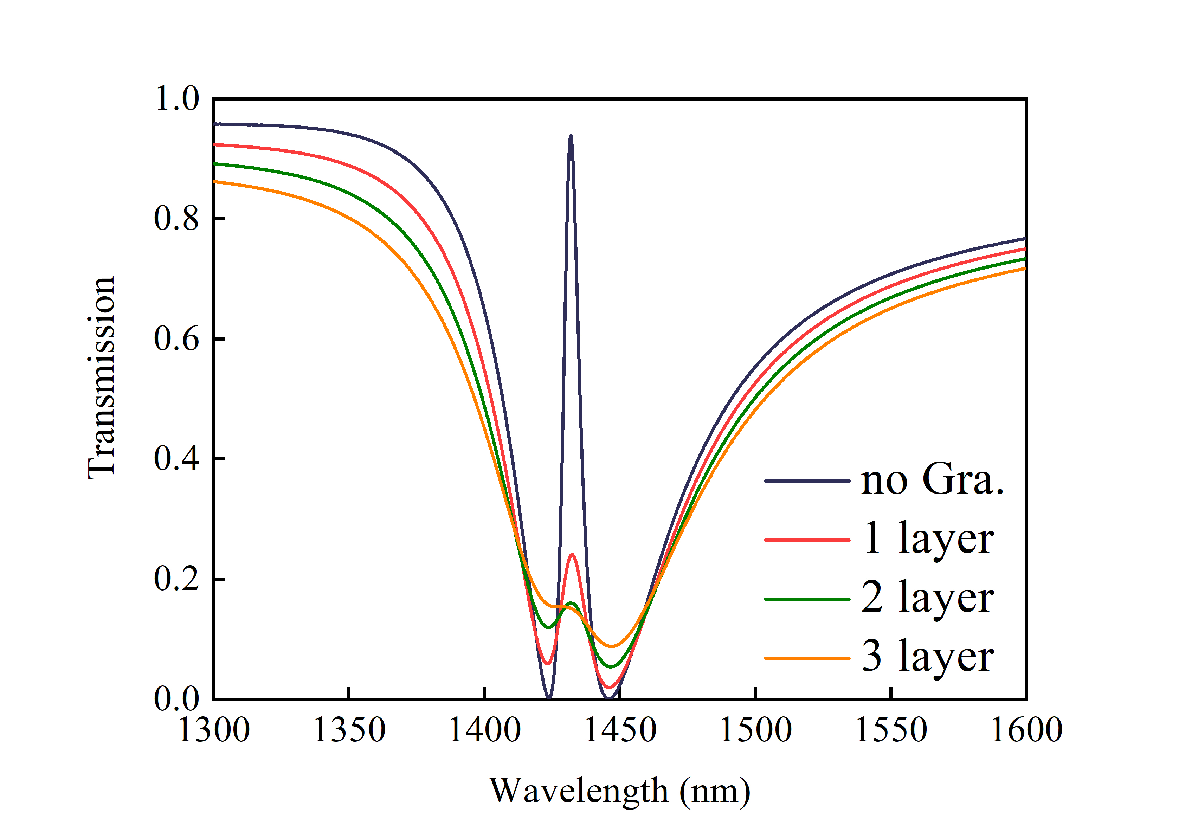}
\caption{The simulated transmission spectra of the hybrid dielectric-graphene metamaterials via altering the layer number.\label{fig6}}
\end{figure}

Under practical conditions, there are inevitable impurities generated during the synthesis or the transfer processes of the monolayer graphene, and the multilayer graphene would probably exist in stacked configurations where half the atoms in one layer lie on half the atoms in others. Therefore, we finally discuss the influence of the layer number of graphene on the EIT resonance. Previous works have demonstrated that the stacked multilayer graphene still behaves like the monolayer graphene due to the electrical decoupling and the conductivity becomes proportional to the layer number\cite{hass2008multilayer,Wang2019}. To obtain maximum possible modulation depth, the Fermi level is fixed to 0 eV (undoped graphene).  Figure. \ref{fig6} shows the simulated transmission spectra, we can see the presence of the multilayer graphene further reduces and even switches off the transparency window. The transmission amplitude decreases to 0.24, 0.16 and 0.15 via altering the layer number from 1 to 3, and the maximum modulation depth can be calculated as $\Delta T=81\%$ with the trilayer graphene. The numerical results are consistent with the scaling relationship between the total conductivity and the layer number of graphene, and substantiate our proposed physical origin. In addition, the modulation effect here can also be interpreted as the "sensitivity" of the EIT resonance to graphene, which is highly desired in the field of optical sensing. In general, metamaterial near field sensing requires the analyte with hundreds of nanometers in thickness\cite{kabashin2009plasmonic,tobing2013deep,xiao2017ultrasensitive}, however, in the present case, multilayer graphene ($\sim1$ nm) is accurately detected, showing an ultralow thickness threshold of $\lambda/1000$ thinner than free space wavelength, which can be promoted to other 2D materials or biomolecules with the similar conductivity.

\section{Conclusions}\label{sec4}

In conclusions, we propose the hybrid structure consisting of all-dielectric metamaterials loaded with graphene and realize the active control of the EIT resonance in the near infrared regime. The physical origin is well explained with the coupling effects between the in-plane electric field of the all-dielectric metamaterials and graphene. With the manipulation of graphene conductivity via shifting the Fermi level, the light absorption through the interband loss of graphene can be switched on (undoped graphene) and off (heavily doped graphene), thus inducing the remarkably high modulation depth $\Delta T=72\%$ in the transmission amplitude of the EIT resonance. With further scaling graphene conductivity via altering the layer number, the modulation depth can reach the maximum up to $\Delta T=81\%$. Such high modulation performance together with the underlying physical origin can be strategically important in designing active hybrid metadevices in the near infrared, such as spatial light modulators, switches and sensors.

\section*{Acknowledgments}
This work is supported by the National Natural Science Foundation of China (Grant No. 11847132, 11947065 and 61901164), the Natural Science Foundation of Jiangxi Province (Grant No. 20202BAB211007), the Interdisciplinary Innovation Fund of Nanchang University (Grant No. 2019-9166-27060003), the Natural Science Research Project of Guizhou Minzu University (Grant No. GZMU[2019]YB22) and the China Scholarship Council (Grant No. 202008420045). 

The authors would also like to thank Dr. S. Li for her guidance on the effective multipole expansion.


\begin{thebibliography}{10}
	\expandafter\ifx\csname url\endcsname\relax
	\def\url#1{{\tt #1}}\fi
	\expandafter\ifx\csname urlprefix\endcsname\relax\def\urlprefix{URL }\fi
	\providecommand{\eprint}[2][]{\url{#2}}
	
	\bibitem{zheludev2012metamaterials}
	Zheludev N~I and Kivshar Y~S 2012 {\em Nat. Mater.\/} {\bf 11} 917
	
	\bibitem{fedotov2007sharp}
	Fedotov V, Rose M, Prosvirnin S, Papasimakis N and Zheludev N 2007 {\em Phys.
		Rev. Lett.\/} {\bf 99} 147401
	
	\bibitem{zhang2008plasmon}
	Zhang S, Genov D~A, Wang Y, Liu M and Zhang X 2008 {\em Phys. Rev. Lett.\/}
	{\bf 101} 047401
	
	\bibitem{liu2009plasmonic}
	Liu N, Langguth L, Weiss T, K{\"a}stel J, Fleischhauer M, Pfau T and Giessen H
	2009 {\em Nat. Mater.\/} {\bf 8} 758
	
	\bibitem{jahani2016all}
	Jahani S and Jacob Z 2016 {\em Nat. Nanotechnol.\/} {\bf 11} 23
	
	\bibitem{kuznetsov2016optically}
	Kuznetsov A~I, Miroshnichenko A~E, Brongersma M~L, Kivshar Y~S and
	Luk’yanchuk B 2016 {\em Science\/} {\bf 354} aag2472
	
	\bibitem{Baranov2017}
	Baranov D~G, Zuev D~A, Lepeshov S~I, Kotov O~V, Krasnok A~E, Evlyukhin A~B and
	Chichkov B~N 2017 {\em Optica\/} {\bf 4} 814
	
	\bibitem{miroshnichenko2012fano}
	Miroshnichenko A~E and Kivshar Y~S 2012 {\em Nano Lett.\/} {\bf 12} 6459--6463
	
	\bibitem{zhang2013near}
	Zhang J, MacDonald K~F and Zheludev N~I 2013 {\em Opt. Express\/} {\bf 21}
	26721--26728
	
	\bibitem{zhang2014electromagnetically}
	Zhang J, Liu W, Yuan X and Qin S 2014 {\em J. Opt.\/} {\bf 16} 125102
	
	\bibitem{yang2014all}
	Yang Y, Kravchenko I~I, Briggs D~P and Valentine J 2014 {\em Nat. Commun.\/}
	{\bf 5} 5753
	
	\bibitem{hopkins2015interplay}
	Hopkins B, Filonov D~S, Miroshnichenko A~E, Monticone F, Alu A and Kivshar Y~S
	2015 {\em ACS Photonics\/} {\bf 2} 724--729
	
	\bibitem{liu2017high}
	Liu Z, Fu G, Liu X, Liu Y, Tang L, Liu Z and Liu G 2017 {\em J. Phys. D: Appl.
		Phys.\/} {\bf 50} 165106
	
	\bibitem{tuz2018high}
	Tuz V~R, Khardikov V~V, Kupriianov A~S, Domina K~L, Xu S, Wang H and Sun H~B
	2018 {\em Opt. Express\/} {\bf 26} 2905--2916
	
	\bibitem{Li2019}
	Li S, Zhou C, Liu T and Xiao S 2019 {\em Phys. Rev. A\/} {\bf 100}
	
	\bibitem{Zhang2019}
	Zhang Z, Yang Q, Gong M and Long Z 2019 {\em J. Phys. D: Appl. Phys.\/} {\bf
		53} 075106
	
	\bibitem{gu2012active}
	Gu J, Singh R, Liu X, Zhang X, Ma Y, Zhang S, Maier S~A, Tian Z, Azad A~K, Chen
	H~T {\em et~al.\/} 2012 {\em Nat. Commun.\/} {\bf 3} 1151
	
	\bibitem{xu2016frequency}
	Xu Q, Su X, Ouyang C, Xu N, Cao W, Zhang Y, Li Q, Hu C, Gu J, Tian Z {\em
		et~al.\/} 2016 {\em Opt. Lett.\/} {\bf 41} 4562--4565
	
	\bibitem{fan2017electromagnetic}
	Fan Y, Qiao T, Zhang F, Fu Q, Dong J, Kong B and Li H 2017 {\em Sci. Rep.\/}
	{\bf 7} 40441
	
	\bibitem{manjappa2017hybrid}
	Manjappa M, Srivastava Y~K, Solanki A, Kumar A, Sum T~C and Singh R 2017 {\em
		Adv. Mater.\/} {\bf 29} 1605881
	
	\bibitem{ahmadivand2017active}
	Ahmadivand A, Gerislioglu B and Pala N 2017 {\em J. Phys. Chem. C\/} {\bf 121}
	19966--19974
	
	\bibitem{zhu2018controlling}
	Zhu W, Yang R, Fan Y, Fu Q, Wu H, Zhang P, Shen N~H and Zhang F 2018 {\em
		Nanoscale\/} {\bf 10} 12054--12061
	
	\bibitem{zhao2016graphene}
	Zhao X, Yuan C, Zhu L and Yao J 2016 {\em Nanoscale\/} {\bf 8} 15273--15280
	
	\bibitem{he2017implementation}
	He X, Yang X, Lu G, Yang W, Wu F, Yu Z and Jiang J 2017 {\em Carbon\/} {\bf
		123} 668--675
	
	\bibitem{he2018graphene}
	He X, Liu F, Lin F and Shi W 2018 {\em Opt. Express\/} {\bf 26} 9931--9944
	
	\bibitem{xia2018plasmonically}
	Xia S~X, Zhai X, Wang L~L and Wen S~C 2018 {\em Photonics Res.\/} {\bf 6}
	692--702
	
	\bibitem{Jia2019}
	Jia W, Ren P, Jia Y and Fan C 2019 {\em J. Phys. Chem. C\/} {\bf 123}
	18560--18564
	
	\bibitem{Guan2020}
	Guan J, Xia S, Zhang Z, Wu J, Meng H, Yue J, Zhai X, Wang L and Wen S 2020 {\em
		Nanoscale Res. Lett.\/} {\bf 15}
	
	\bibitem{li2016monolayer}
	Li Q, Cong L, Singh R, Xu N, Cao W, Zhang X, Tian Z, Du L, Han J and Zhang W
	2016 {\em Nanoscale\/} {\bf 8} 17278--17284
	
	\bibitem{xiao2017strong}
	Xiao S, Wang T, Jiang X, Yan X, Cheng L, Wang B and Xu C 2017 {\em J. Phys. D:
		Appl. Phys.\/} {\bf 50} 195101
	
	\bibitem{chen2017study}
	Chen X and Fan W 2017 {\em Opt. Lett.\/} {\bf 42} 2034--2037
	
	\bibitem{xiao2018active}
	Xiao S, Wang T, Liu T, Yan X, Li Z and Xu C 2018 {\em Carbon\/} {\bf 126}
	271--278
	
	\bibitem{Hong2019}
	Hong Q, Luo J, Wen C, Zhang J, Zhu Z, Qin S and Yuan X 2019 {\em Opt.
		Express\/} {\bf 27} 35914
	
	\bibitem{Xia2020}
	Xia Y, Wang J, Zhang Y, Shan Y, Dai Y, Chen A, Shen T, Wu S, Liu X, Shi L and
	Zi J 2020 {\em Adv. Opt. Mater.\/}  2000264
	
	\bibitem{argyropoulos2015enhanced}
	Argyropoulos C 2015 {\em Opt. Express\/} {\bf 23} 23787--23797
	
	\bibitem{liu2017toroidal}
	Liu G~D, Zhai X, Xia S~X, Lin Q, Zhao C~J and Wang L~L 2017 {\em Opt.
		Express\/} {\bf 25} 26045--26054
	
	\bibitem{Xiao2019}
	Xiao S, Liu T, Zhou C, Jiang X, Cheng L, Liu Y and Li Z 2019 {\em J. Phys. D:
		Appl. Phys.\/} {\bf 52} 385102
	
	\bibitem{Sun2020}
	Sun G, Peng S, Zhang X and Zhu Y 2020 {\em Nanomaterials\/} {\bf 10} 1064
	
	\bibitem{palik1985handbook}
	Palik E and Ghosh G 1985 {\em Handbook of Optical Constants of Solids (Orlando,
		FL: Academic)\/}
	
	\bibitem{zhang2015towards}
	Zhang J, Zhu Z, Liu W, Yuan X and Qin S 2015 {\em Nanoscale\/} {\bf 7}
	13530--13536
	
	\bibitem{xiao2016tunable}
	Xiao S, Wang T, Liu Y, Xu C, Han X and Yan X 2016 {\em Phys. Chem. Chem.
		Phys.\/} {\bf 18} 26661--26669
	
	\bibitem{hass2008multilayer}
	Hass J, Varchon F, Millan-Otoya J~E, Sprinkle M, Sharma N, de~Heer W~A, Berger
	C, First P~N, Magaud L and Conrad E~H 2008 {\em Phys. Rev. Lett.\/} {\bf 100}
	125504
	
	\bibitem{Wang2019}
	Wang J, Chen A, Zhang Y, Zeng J, Zhang Y, Liu X, Shi L and Zi J 2019 {\em Phys.
		Rev. B\/} {\bf 100} 075407
	
	\bibitem{kabashin2009plasmonic}
	Kabashin A, Evans P, Pastkovsky S, Hendren W, Wurtz G, Atkinson R, Pollard R,
	Podolskiy V and Zayats A 2009 {\em Nat. Mater.\/} {\bf 8} 867
	
	\bibitem{tobing2013deep}
	Tobing L~Y, Tjahjana L, Zhang D~H, Zhang Q and Xiong Q 2013 {\em Sci. Rep.\/}
	{\bf 3} 2437
	
	\bibitem{xiao2017ultrasensitive}
	Xiao S, Wang T, Liu Y, Han X and Yan X 2017 {\em Plasmonics\/} {\bf 12}
	185--191
	
\end{thebibliography}

\section*{References}
\providecommand{\newblock}{}

\end{document}